\documentclass[twoside,onecolumn,a4paper]{article}
\usepackage{amsmath, amsthm, amssymb}
\usepackage[super,sort&compress,comma]{natbib} 
\usepackage{times}

\usepackage{graphicx} 
\usepackage{hyperref}

\begin{document}

\thispagestyle{plain}
\renewcommand{\thefootnote}{\fnsymbol{footnote}}
\renewcommand\footnoterule{\vspace*{1pt}%
\hrule width 3.4in height 0.4pt \vspace*{5pt}} 
\setcounter{secnumdepth}{5}

\makeatletter 
\def\subsubsection{\@startsection{subsubsection}{3}{10pt}{-1.25ex plus -1ex minus -.1ex}{0ex plus 0ex}{\normalsize\bf}} 
\def\paragraph{\@startsection{paragraph}{4}{10pt}{-1.25ex plus -1ex minus -.1ex}{0ex plus 0ex}{\normalsize\textit}} 
\renewcommand\@biblabel[1]{#1}            
\renewcommand\@makefntext[1]%
{\noindent\makebox[0pt][r]{\@thefnmark\,}#1}
\makeatother 
\renewcommand{\figurename}{\small{Fig.}~}

\setlength{\arrayrulewidth}{1pt}
\setlength{\columnsep}{6.5mm}
\setlength\bibsep{1pt}

\noindent\LARGE{\textbf{Water formation at low temperatures by surface O$_2$ hydrogenation II: the reaction network}}
\vspace{0.6cm}

\noindent\large{\textbf{H.~M.~Cuppen,$^{\ast}$\textit{$^{a,b}$}, S.~Ioppolo\textit{$^{a}$}, C.~Romanzin\textit{$^{a\ddag}$} and H.~Linnartz\textit{$^{a}$}}}\vspace{0.5cm}


\noindent \textbf{\small{DOI: 10.1039/C0CP00251H}}
\vspace{0.6cm}

\noindent \normalsize{Water is abundantly present in the Universe. It is the main component of interstellar ice mantles and a key ingredient for life. Water in space is mainly formed through surface reactions. Three formation routes have been proposed in the past: hydrogenation of surface O, O$_2$, and O$_3$. In a previous paper [Ioppolo \textit{et al., Astrophys.~J}, 2008, \textbf{686}, 1474] we discussed an unexpected non-standard zeroth order H$_{2}$O$_{2}$ production behaviour in O$_2$ hydrogenation experiments, which suggests that the proposed reaction network is not complete, and that the reaction channels are probably more interconnected than previously thought. In this paper we aim to derive the full reaction scheme for O$_2$ surface hydrogenation and to constrain the rates of the individual reactions. This is achieved through simultaneous H-atom and O$_2$ deposition under ultra-high vacuum conditions for astronomically relevant temperatures. Different H/O$_2$ ratios are used to trace different stages in the hydrogenation network. The chemical changes in the forming ice are followed  by means of Reflection Absorption Infrared Spectroscopy (RAIRS). New reaction paths are revealed as compared to previous experiments. Several reaction steps prove to be much more efficient (H+O$_2$) or less efficient (H+OH and H$_2$+OH) than originally thought. These are the main conclusions of this work and the extended network concluded here will have profound implications for models that describe the formation of water in space.
}
\vspace{0.5cm}

\includegraphics[width=.45\textwidth]{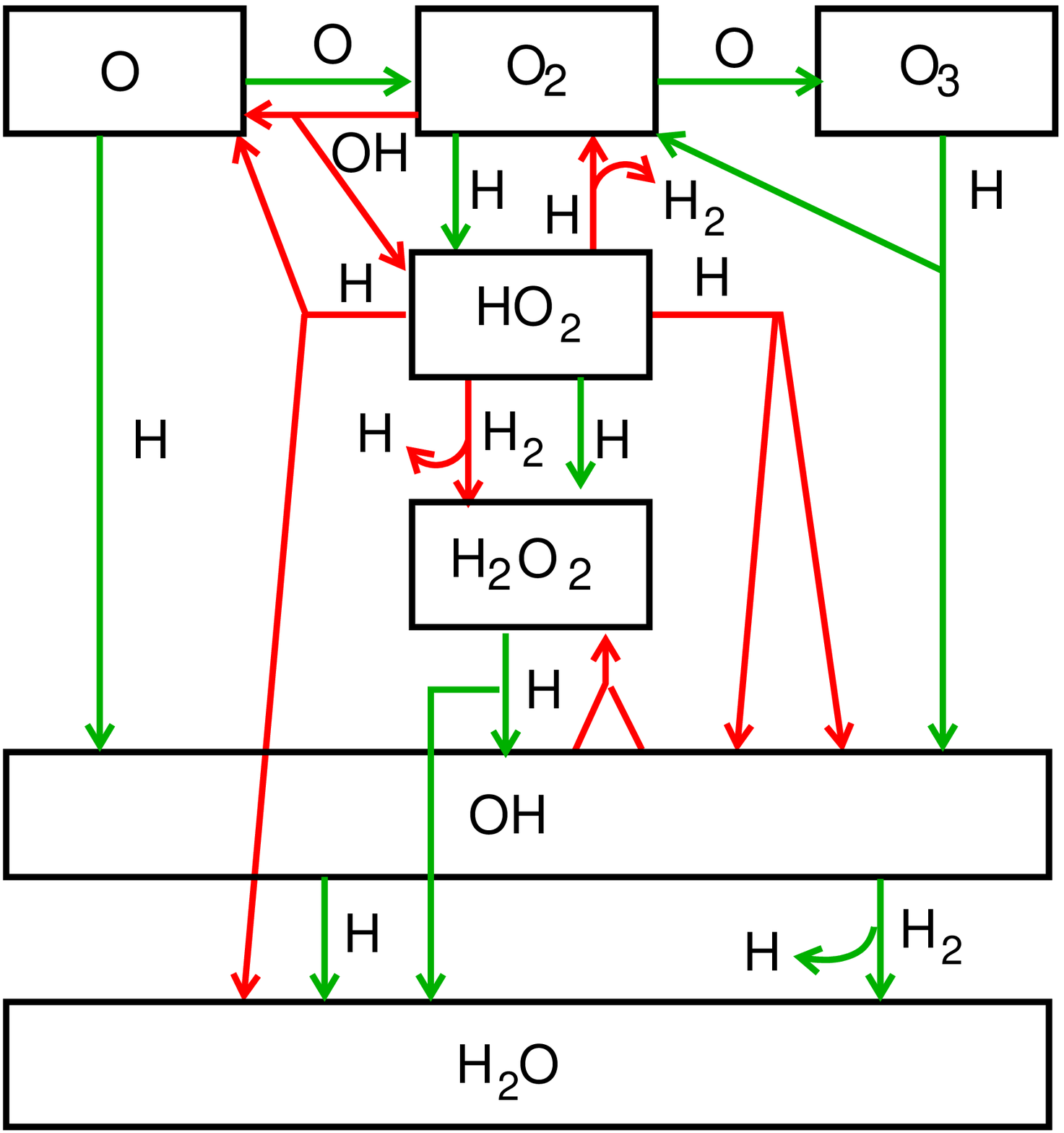}

\noindent
Laboratory experiments show that the formation of water in space is much more complex (green + red arrows) than previously thought (just green). 

\vspace{2ex}
\noindent
Due to a copyright agreement we are not allowed to publish the full paper on  arXiv.org. Please look \href{http://dx.doi.org/10.1039/C0CP00251H}{here} for the paper. We apologise for any inconvenience.

\footnotetext{\textit{$^{a}$} Sackler Laboratory for Astrophysics, Leiden Observatory, Leiden University, P. O. Box 9513, 2300 RA Leiden, The Netherlands}
\footnotetext{\textit{$^{b}$} Leiden Observatory, Leiden University, P. O. Box 9513, 2300 RA Leiden, The Netherlands}
\footnotetext{\dag~Present address: LPMAA, Universit\'e Pierre et Marie Curie, Paris}

\end{document}